\pdfoutput=1
\documentclass[12pt,journal,compsoc]{IEEEtran}
\usepackage{cite}
\usepackage{graphicx}
\usepackage{hyperref} 
\usepackage{amsmath}

\begin{document}
\title{SEAL's operating manual:\\ a Spatially-bounded Economic Agent-based Lab}
\author{Bernardo~Alves~Furtado, Isaque~Daniel~Rocha~Eberhardt, and~Alexandre~Messa
\IEEEcompsocitemizethanks{\IEEEcompsocthanksitem B. Furtado, I. Eberhardt and A. Silva are with the Institute for Applied Economic Research, Government of Brazil, Bras\'ilia, Brazil. B. Furtado is also a researcher at CNPq\protect\\
E-mail: bernardo.furtado@ipea.gov.br}}

\markboth{Research Report}
{Shell \MakeLowercase{\textit{}}}

\IEEEtitleabstractindextext{%
\begin{abstract}
This text reports in detail how SEAL, a modeling framework for the economy based on individual agents and firms, works. Thus, it aims to be an usage manual for those wishing to use SEAL or SEAL's results. As a reference work, theoretical and research studies are only cited. SEAL is thought as a Lab that enables the simulation of the economy with spatially bounded microeconomic-based computational agents. Part of the novelty of SEAL comes from the possibility of simulating the economy in space and the instantiation of different public offices, i.e. government institutions, with embedded markets and actual data. SEAL is designed for Public Policy analysis, specifically those related to Public Finance, Taxes and Real Estate.
\end{abstract}

\begin{IEEEkeywords}
Agent-based modeling, public policy, Python, computational economics, municipalities, public finance.
\end{IEEEkeywords}}

\maketitle

\section{Introduction}
\IEEEPARstart{T}{his} text reports in detail how SEAL works.\footnote{This applies to SEAL version 2.0 as of September 2016.} Thus, it aims to be a  a reference  for those wishing to use SEAL as a model, use SEAL's outputs, study its details, or work as a SEAL's collaborator. As a manual, the focus of the text is not on a specific research question or a literature review. Thus, the accompanying literature is not discussed in the report. We direct the interested reader to other sources \cite{macal_everything_2016,wilensky_introduction_2015,epstein_generative_2011,miller_complex_2007,epstein_growing_1996,tesfatsion_agent-based_2006} Rather, here we describe the framework, the Lab, that we hope may be used for a number of applications. Henceforth, we have applied it to \cite{furtado_simple_2016,furtado_da_2016}. This report follows the ODD protocol recommendations of Grimm and colleagues \cite{grimm_odd_2010}. However, we aim at being more specific, describing in detail the components of the model.

We refer to SEAL as a Lab or a framework that enables the simulation of the economy, its markets, considering spatially bounded computational agents. SEAL simulates the economy in space, its markets and government institutions. SEAL is specifically designed to study Public Policy, with an emphasis on Public Finance, Taxes and Real Estate.

This claim of usability derives directly from the available modeling elements present in the model. The Lab comprises Citizens, organized as families, living and moving among households, traveling to firms – both of which: households and firms, are spatially geocoded within actual municipalities' boundaries. Citizens are born, age and die and they are initiated in the model according to official statistics. Municipalities' government is also present. Finally, interaction amongst agents happens in three markets: labor, real estate and goods. All those agents are operated in an agent-based model simulated in \texttt{Python 3.4.4}.

Besides this introduction, this report includes a simple statement of purpose (\autoref{section 2}), a description of the process overview and scheduling (\autoref{section 3}), a full description of the agents, its methods, the markets (\autoref{section 4}) and the different possibilities of simulations to run (\autoref{section 5}). Section 6 describes all the necessary data input, parameters setting and demographic and spatial information necessary to run SEAL for different configurations (\autoref{section 6}). Section 7 delineates the outputs of SEAL (\autoref{section 7}). This report concludes (\autoref{section 8}) with the description of the design concepts, following \cite{grimm_odd_2010}, and the possibilities we see for this framework (\autoref{section 9}).

\section{Purpose}\label{section 2}

SEAL was designed specifically to tackle public policy concerns. A main justification can be found in \cite{furtado_modeling_2015}. As such, SEAL provides an economic-based framework to answer a number of questions. A proposed list is available in \cite{furtado_simple_2016}, section 4. Illustratively, we could cite: general evaluation of tax impacts; economic impacts on mobility; general impacts of agents' decisions; impacts of changes in government spatial outreach, to name a few.

\section{Process overview and schedule}\label{section 3}

\begin{figure*}
	\includegraphics[width=1\textwidth, height=.88\textheight]{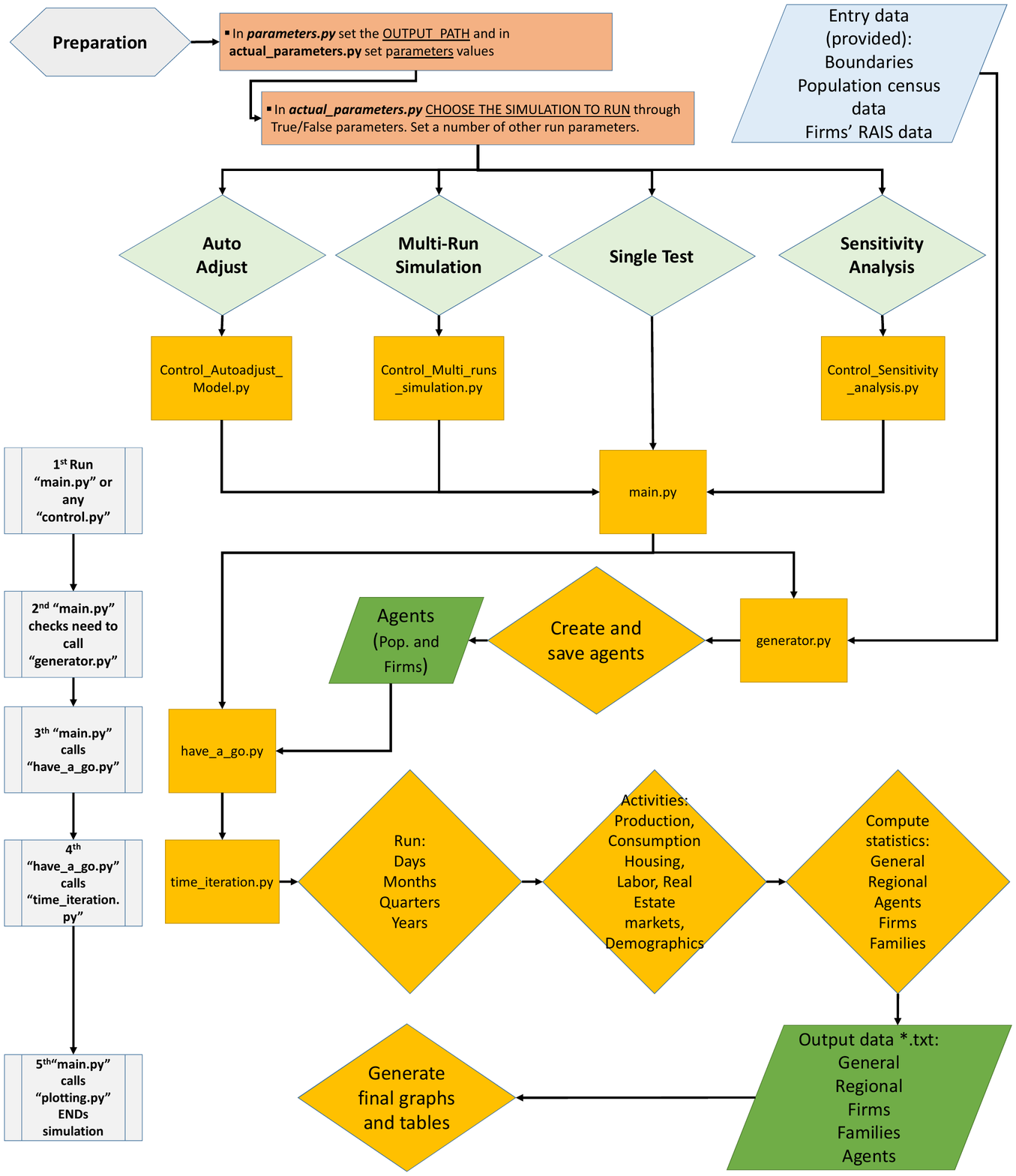}
	\centering
	\caption{SEAL flowchart for the major processes and \texttt{python} modules. The grey objects are definitions or data to start running a simulation. Population data come from Brazilian demographics census; number and size of firms from RAIS. The orange objects represent decisions to be made while running the model. Rectangles are parameters definitions and the lozenges represent the kind of simulation chosen. The gold rectangles represent the most important modules, which control the model and run the processes. Green parallelograms represent output data (graphs, maps, txt and csv files).}
	\label{fig_sim}
\end{figure*}

SEAL is setup so that the modeler can run a single test or multiple runs, in order to make sure the results are consistent (see \autoref{section 5}). When running a single run, the modeler should simply type:
\texttt{python main.py} in a \texttt{python 3.X} interpreter (\autoref{fig_sim}). However, it would be wise to set a number of parameters first (see \autoref{section 6}) and make sure all the necessary libraries are installed.

\subsection{\texttt{Time\_iteration} module}
SEAL is organized around a discrete time frame. As such, time is divided into days, months, quarters and years. It is possible to insert periodical activities, for any of the agents in the model, at any one of these periods. 

The first function that runs before the actual \texttt{time\_iteration} is \texttt{have\_a\_go}. 

\texttt{Have\_a\_go} initiates \texttt{my\_simulation} as an instance of \texttt{time\_iteration.TimeControl} class. Then it runs up to the total number of days setup for the full simulation. Typically, since we only consider working days, months have 21 days, thus, we run for 5.040 days (which is equivalent to 20 years).

\subsubsection{Daily activities}
Daily activities are restricted to production updates. On \texttt{Day 0} of the simulation, however, some setup is put in place before actual simulation begins. This includes the single time a product is created \footnote{The structure of the Lab is compatible with the idea that companies may create more products with different characteristics during simulation. The current configuration does not differentiate among products, which are homogeneous.}  and a first round of labor market matching up, so that firms may have employees and be able to produce during the first month of simulation.

\subsubsection{Monthly activities}
In the beginning of every month, the demographics dynamic is updated. Given the citizen's month of birth, the probabilities of her death and of her giving birth (see details at section 4.7) are calculated. Then, firms pay salaries (workers receive their salaries before they can go shopping).

Families distribute their money equally among their members. Then, each member acquires his consumption goods, and immediately consumes them (see details at goods market (see section 4.6.1). Governments collect taxes from consumption. In turn, firms update or maintain their prices according to their inventory levels (see section 4.5). Finally, each government municipality spends the collected taxes on public goods. The region quality of life (QLI) is improved by the same amount as the government spending per capita. However, a \texttt{TREASURE\_INTO\_SERVICES} parameter adjusts the numbers so that QLIs remain close to a 0 to 1 indicator.

Before the end of the month, the labor market (see section 4.6.1) is processes, followed by one run of the real estate market (see section 4.6.3). Finally, statistics and outputs are calculated and saved (section 7). Presently, quarters and years are used exclusively for statistics computation.  

Four other functions are located at \texttt{time\_iteration} for importing or usage reasons. However, they pertain to the labor market and the equalization of resources among family members. 

In short, activities run every month are subsequently:
\begin{enumerate}
	\item Record month
	\item Process demographics (aging, births, deaths).
	\item Firms make payments.
	\item Families redistribute their own cash within members.
	\item Family members consume (Goods market).
	\item Governments collect taxes.
	\item Governments spend the collected taxes on life quality improvement (QLI).
	\item Firms calculate profits and update prices.
	\item Labor market is processed.
	\item Real estate market is processed.
	\item Statistics and output are processed.
\end{enumerate}

\section{Actual entities and processes}\label{section 4}

This section aims at detailing each model agent variables and methods. All five following agent types are treated as classes and instantiated once (see section 6.3) before running the simulation.

\subsection{Citizens}

Agents have the following variables: 

\begin{itemize}
	\item fixed: \texttt{id, gender, month of birth, qualification, family\_id}. 
	\item variable: \texttt{age, money} (amount owned at any given moment), \texttt{savings, firm\_id, utility} (an indicator of cumulative amount consumed), \texttt{address (osgeo, ogr), distance, region\_id}.
\end{itemize}

When instantiated, agents are given an id (in successive order of generation), a gender, an initial age, a fixed qualification and some initial money. The number of agents, their gender and their age come from official data (see section 6.3). Qualification is drawn from a gamma distribution with $\alpha=\beta=3$. Thus, it is independent of age and fixed throughout the model. Money is drawn from a uniform distribution in (20, 40). 

Agents have a number of methods. The usual \texttt{get()} methods are used to return variables values, to check conditions related to age (such as \texttt{is\_minor} and \texttt{is\_retired}), and update variables (such as \texttt{update\_age} and \texttt{update\_money} – given a certain amount). 

A number of methods are related to family business. \texttt{set\_family}, which then never changes, \texttt{get\_family\_id} and a check to see if the agent \texttt{belongs\_to\_family}. Address and consequent region membership change according to the changes the family does. Every time a family changes addresses, values are updated at the agent level. 

Methods related to the workplace include setting and getting workplace and a check to see if the agent \texttt{is\_employed}. When \texttt{firm\_id is None}, then either agent is unemployed or is out of working age. The variable distance always refers to the distance between current household where the agent resides and current firm, when employed. Otherwise, it is set to \texttt{None}. 

Each agent also has a method that calculates distance from his or her current house location to prospective and current place of employment. 
Information that is printed and outputted in TXT format individually for each agent includes: \texttt{id, gender, qualification, age, money, firm\_id, utility, address, and region\_id}. 

\subsection{Citizens' families}

Families have a single \texttt{id} variable. They also hold values – that are constantly updated – for \texttt{balance, savings, household\_id} and \texttt{region\_id} (initiated at \texttt{None}), \texttt{house\_price, address} and \texttt{house}. House is a variable that actually contains the full instance of a household. It is useful to access the house price, its address and region. This scheme is necessary because families move from house to house, given the endogenous changes of the model as a whole. 

Finally, an important variable of families is an inventory \texttt{members} (actually, a Python dictionary) that contains all the instances of the families' members. Related proper methods are \texttt{is\_member} and \texttt{num\_members}, which report the family size.

The methods in Families include typical get and set ones; an \texttt{add\_agent}, which brings the agent instance into members and sets the \texttt{family\_id} inside the agent instance. A counterpart method is that of \texttt{death\_member}, which removes the member from the family (at the death event). Families also have access to \texttt{sum\_balance, sum\_savings and update\_balance}, which is a method that divides a received amount equally among family members.

Other two important pair of methods are \texttt{assign\_house} which updates \texttt{household\_id, region\_id}, the \texttt{house} itself, and \texttt{empty\_house} (which makes all those variables into \texttt{None}, and also calls \texttt{empty()} within the house class). \texttt{set\_address} updates address and \texttt{region\_id} for all members within the family.

A relevant function \texttt{consume()} is hosted at the level of the family, but it will be described in the Goods and Services Market, section 4.6.1.

Finally, families calculate internally the proportion of employed members as well as average utility.

\subsection{Households}

Households are simple, fixed class objects that contain mostly get and set methods. Fixed variables include: \texttt{house\_id, address, size, region\_id, quality}, all of which are set at the beginning via the generator module (section 6.3). Among those methods, we can list: \texttt{get\_family\_id, update\_ownership, is\_occupied, add\_family, and empty} – which changes \texttt{family\_id} into \texttt{None}.

Non-fixed variables include \texttt{price, family\_id and ownership} – so that both the current family living on a given household as well as its owner can change dynamically. 

These processes of changing ownership and residence both happen at the real estate market (section 4.6.3). \texttt{Price} is updated through the relation \texttt{size} times \texttt{quality} times a region Quality of Life Index (the latter is updated at the \texttt{Government} module, according to the amount of taxes collected).

\subsection{Government}

This class represents the individual administrative-political unit – basically, the municipalities. It is generated based on shapefile objects (that contain the geometric features of the boundaries), using \texttt{OSGEO} objects. 

It is the REGION where taxes are applied and the REGION where taxes collected are spent in order to increase the quality of life. The class also contains self-calculated information including firms and citizens within its territory.

In detail. 

The generation of the \texttt{Govern} class object needs to be read as \texttt{ogr from OSGEO}. This implies that an \texttt{address\_envelope}, which is read from \texttt{geometry().GetEnvelope()} is possible as an \texttt{\_\_init\_\_ method}. Addresses also comes from \texttt{geometry()}.

Apart from the geometric features, Govern has a name – which is extracted from the available fields associated in the shapefile using \texttt{GetFieldAsString(0)} – and an ID – which also comes from a field \texttt{GetField(0)}. 

Other variables include \texttt{index}, which is the current value of the Quality of Life Index \footnote{For simulation purposes, this \texttt{index} is initiated as the value of the municipal Human Development Index in 2000}, \texttt{treasure}, the current amount of available cash, the \texttt{region\_gdp}, which is set at \texttt{statistics} (section 7.1), its population (pop), and a calculated \texttt{total\_commute}.

The method to calculate the population receives all the families as input and for those families in which \texttt{region\_id} is the same of the current region, the members of such family are added up.

The government spending (that is, the amount of collected taxes) is directly converted in increase in the quality of life index. In order to keep values nearly compatible with empirical data, \texttt{treasure} is normalized dividing it by population, and multiplying the result by a reducing value of \texttt{TREASURE\_INTO\_SERVICES = 0.0005}. 

\subsection{Firms}

Firms are generated at the start of the simulation and remain fixed afterwards. Thus, \texttt{firm\_id, address and region\_id} are fixed variables. Besides, at the beginning of the simulation, each firm has a small amount of cash that is stored at \texttt{total\_balance}.

However, the number of employed workers, their quality, and the price of the firms' product vary through time. Variables used to follow the firms dynamics include: \texttt{last\_qtr\_balance}, updated every three months; \texttt{profit} \footnote{\texttt{Profit} starts at 1 in order to guarantee an opening positive value.} whether current \texttt{total\_balance} is above \texttt{last\_qtr\_balance $-$ amount\_sold} that saves the cumulative number of products sold; and \texttt{amount\_produced}. The usual get and set methods allow access to the values of the firms' variables.

The firms also contain a dictionary of \texttt{employees} with all of its employees at any given time. As an object oriented programming paradigm, the dictionary carries the full instances of the employed agents, thus providing access to all of the agents' information.
 
Prices can go up, down or remain the same. Prices remain unchanged when either the number of employees or the inventory are zero. 

When a firm’s inventory ($Q$) is above a given parameter ($K$) (\texttt{QUANTITY\_TO\_CHANGE\_PRICES}), then the price of its product is reduced in proportion to the \texttt{MARKUP} parameter. The opposite happens when quantity goes below that same level, and they increase proportional to the \texttt{MARKUP} value.

\begin{equation}
price_t = \left\{
\begin{array}{@{}ll@{}}
price_{t-1} * (1 + markup), & \text{if}\ Q < K\\
price_{t-1} * (1 - markup), & \text{otherwise}
\end{array} \right.
\end{equation}

\texttt{Sale} is the method that operationalizes the goods market. It is detailed in section 4.6.1.

Other methods pertaining the functioning of the firm are related to employees' business. Those include \texttt{add\_employee}, which not only includes the agent in the employee dictionary, but also passes the \texttt{firm\_id} to the \texttt{agent} (who can then store it and thus change his status to employed). The \texttt{obit} method deletes the employee from the firms' dictionary when he or she passes (see \texttt{demographics} module). When \texttt{fire} is called, the employee is also removed from the dictionary of employees of the firm and his \texttt{firm\_id} is set back to \texttt{None}.

A relevant role of the firm is to control staff at all times. 

A method to \texttt{add\_employee} makes sure to register the agent within a firm's dictionary employees. It also adds \texttt{firm\_id} to the agent using method \texttt{set\_workplace}. When an employee dies and is withdrawn from the simulation, he is deleted from the firm's team with \texttt{obit}. When called, \texttt{fire} chooses a (single) random employee and deletes him or her from the staff and also sets workplace to \texttt{None}. Other methods related to employment include a check, \texttt{is\_worker} and \texttt{num\_employees} (which returns the size of current staff). 

Payment of employees is such that when the firm is having profits and cashflow is positive, profits are distributed as salary/bonuses using individual productivity criteria. Otherwise, salary is based on productivity times a unitary price.
 
Payment of employees is calculated in a two-stage process. When the firm is distributing profits, a \texttt{get\_wage\_base} is defined, such that the base salary is the unit plus the percentage of profits over current available cash. So that if the company has a current cashflow of 1,000 and profit was 200, then base salary is incremented by a 20\% margin at 1.20. Otherwise, base salary remains at unit. Then, payment is made exactly according to production. That is, each employee produces the quantity equivalent to qualification exponentiated by a productivity parameter $\alpha$.
 
Thus, salary is the unit (or unit plus profit distribution margin) times days of production (21) times qualification exponentiated by $\alpha$.

\begin{equation}
salary_t = \frac{profit_{t-1}}{cashflow_{t-1}} * qualification^\alpha
\end{equation}

\subsubsection{Products}

Another dictionary is \texttt{inventory} – which contains the list of the firm's products. At the current implementation, there is only a single product. However, code is implemented so that other products could be easily added. In order to save and control the IDs of possible products, there is also a \texttt{product\_index} variable, which starts at 0. 

A special method, \texttt{create\_product}, is used only once, before the simulation actually begins. Products contain information on \texttt{price}, originally set at 1, and \texttt{quantity}, obviously, initially set at 0. They are also produced as a class, stored at \texttt{products} module.
 
\texttt{update\_product\_quantity} is actually the method that implements the production function. Production is based on the quantity and quality of the employees. Given that any employee has a given level of qualification, the quantity produced is each employee's qualification raised to an alpha parameter that indicates the productivity level. Thus:

\begin{equation}
Q_{it} = \sum_{j = 1}^{J}j_i^\alpha
\end{equation}

where $Q_i$ represents the quantity produced by firm $i, j_i$, each employee qualification, and $\alpha$, the productivity parameter. Production only occurs if the firm has a positive number of employees, positive monetary funds and at least one registered product at inventory.

\subsection{Markets}

There are three markets at the current implementation of the model. The operation of goods and services is distributed among methods within the agents and the firms, and is monthly called for every agent at \texttt{time\_iteration}. In turn, the labor market has a module of its own, named communications – indeed, that market was thought out as an announcement board where companies and potential employees meet. Finally, the real estate market also has a module of its own, called \texttt{housing\_market}. 

\subsubsection{Goods and services}

The function \texttt{consume} is located within each family with the following steps. First, the family needs to have a positive amount of cash in order to enter the market. If that amount is below the unit, a random number between 0 and the available amount is chosen to be spent. 

If, on the other hand, that amount is above the unit, the chosen consumption is a percentage of the available amount of money ($M$). Such percentage is given by a number drawn from a beta distribution, so that the average is given by the modeler chosen $\beta$ parameter, i.e. an average propensity to consume

\begin{equation}
  to\_spend \sim \left\{
  \begin{array}{@{}ll@{}}
  U(0, 1) * M, & \text{if}\ M < 1\\
  Beta(1, \frac{1 - \beta}{\beta}) * M, & \text{otherwise}
  \end{array} \right.
\end{equation}

Once the decision of the amount to be spent is made, the agent checks another parameter called \texttt{SIZE\_MARKET}. This parameter determines the number of firms the agent makes contact before doing the consumption decision. 

Third, among the contacted firms, the agent chooses to buy either from the cheapest firm, or the closest one. This decision is random and equally weighted with probability 0.5. 

Once those decisions have been made, consumption is straightforward, given that products are homogeneous. 

If the firm can sell the full agent demand, according to his full amount to spend, the sale occurs. Otherwise, the firm sells as much quantity as it can provide, and returns monetary change for the quantity it cannot provide. 

At each month, the cash that the agent decides not to spend goes to a savings account. This account enables the family to enter the real estate market. 

\texttt{Sales} are located within the firm agents. A simple check is made to see whether the amount available from the consumer is positive. Given the original structure that allowed for more than one product, the amount to spend is distributed equally among the products available. However, for this implementation, there is a single product available and the entire amount is directed to such a product.
 
The first operation within the firm is to transform the cash amount into product quantity, given current price. The resulting quantity is deduced from the firm's inventory and taxes are deduced from the amount paid. The firm's balance is increased by the amount bought, and the treasure of the region where the firm is located is increased by the corresponding tax collected (that is, given the tax rate and the amount spent). The firm registers the amount sold for statistics purposes. 

\subsubsection{Labor}

At the end of every month, the labor market is initiated from \texttt{time\_iteration} that calls \texttt{hire\_fire} and \texttt{look\_for\_jobs}: the first function fills in a list with the firms that are hiring; the second one compiles a list of able and currently unemployed workers. 

The entrance of the firm in the labor market is determined by a parameter (set by the modeler), at the parameters module, called \texttt{LABOUR\_MARKET}. The larger this parameter, less often the firm enters the labor market. If it is set, for instance, at 0.75, it means that, at each month, the firm has a 25\% chance of entering the market – on average, firms would then enter the labor market once every four months. 

At the labor market, the firm decides to hire if its profits are positive (profits are calculated based on cash flow, having last quarter as reference). Otherwise, the firm fires one employee. That is, if the firm has a negative result since last quarter and the strategy is to enter the market, then the firm makes one employee redundant. Decision on the employee is made randomly. That process guarantees a dynamic labor market. 

On the side of likely work seekers, citizens enter the market every month when they fulfill all of the three following conditions: (a) being currently unemployed, (b) being of age 16 or older, (c) being active in the market (up to 69 years old).
 
The module \texttt{communications}, that makes the matching among firms and employees, has a class called \texttt{Posting}. There are two lists, one of candidates and one of available postings. They contain all the firms and citizens. The method that makes the match is \texttt{assign\_post}. 

First, both lists are sorted: the firms offering posts are sorted from those paying the highest base salaries; and the agents looking for jobs are sorted by years of qualification (and, implicitly, productivity). Then, the firm paying the highest salary chooses first: the match is made randomly with a 0.5 probability for the candidate living the closest to the firm or the most qualified in the list. While there are firms offering opened positions and candidates interested, the match goes on. Every month both lists are emptied and the process restarts. 

\subsubsection{Real estate}

The houses prices are updated according to current Quality of Life Indexes (QLI) of their respective regions (municipalities). Thus, every time the municipality collects taxes and spends in public goods, the offering prices of households increase accordingly. In order to make its value comparable to actual Human Development Index for Municipalities, the QLI is normalized by multiplying it by a parameter ($k$) \texttt{TREASURE\_INTO\_SERVICE} (typically set at 0.0005). Further, QLI is weighted by the change in population ($N$).

\begin{equation}
QLI_t = \frac{QLI_{t-1} * N_{t - 1}}{N_t} + \frac{treasure * k}{N_t}
\end{equation}

Once it is empty, the house enters the real estate market. In turn, families entering the market are chosen according to the parameter \texttt{PERCENTAGE\_CHECK\_NEW\_LOCATIONS}, which is a percentage of the universe of families. 

The price of each house is then updated, according to its $size$ and $quality$ and region.

\begin{equation}\label{eq5}
Price_{it} = size * quality * QLI_t
\end{equation}

The lists of houses available and interested families are sorted according, respectively, to house prices and the amount of families' savings. The first step is for the family to acquire a new ownership. Families that own more than one house decides, at the end of the real estate market procedure, whether to move or not. 

Starting from the family with the largest amount, the family will buy the most expensive house it can afford, if any. 

Once the matching family-house has been made, the savings of the buying family is deduced of the price paid and the family who previously owned the house receives the same amount. The actual transfer of registering of house ownership is then made, followed by the removal of the house from the sales list. 

Once the family has successfully accomplished a purchase, it verifies if it is worth moving or not by checking the quality of the houses and the proportion of employed in the house. Thus, there are three options: 
\begin{enumerate}
	\item if the family lives in the best house, but all members of the house are unemployed (or out of the market, such as children and retired members), then the family moves to the second best house; 
	\item if the family does not live in the house with the best quality but at least one person is employed, the family moves into the best house; 
	\item all other options maintain the family in the same house.
\end{enumerate}

Once the decision to move has been made, the family empties the previous house and register the new house, the new address and, eventually, the new region. 

\subsection{Dmographics}

\texttt{Demographics} is the module that controls aging, dying and giving births of agents. It is implemented mainly as one function \texttt{check\_demographics}.
 
The first step is to check the current year of the simulation (by construction, the first year is 2000). All of the citizens age. Then, according to their gender and month of birth, a death probability, derived from official IBGE data estimates, is evaluated and saved. 

Females then evaluate a probability to give birth if they are between 15 and 49 years old. If so, a new agent, child with age zero, is generated and it is set to live within the mother's family.
 
Then, a number is drawn to evaluate the death of agents. When death occurs, agents are withdrawn from the agents list, from their families and from their firms. For statistical purposes, they are added to a new list called \texttt{my\_graveyard}.

\subsection{Spatial boundness}

In the present model, distance is used as one of the criterions for two decisions: when firms choose candidates for employment; and when citizens choose a firm in order to consume its product. 

Distance is also used to compute a regional commuting distance, defined as the sum of the commuting distances of all employed citizens living in a given region.

Furthermore, space plays a role in the model when, at the beginning of the simulation, households, firms and citizens are spatially allocated. This allocation is made considering urban and rural proportion for each municipality (see details at section 6.3).

\section{Running different simulations}\label{section 5}

The model can run in four different ways: 5.1 Simple; 5.2 Sensitivity analysis; 5.3 Multi-run; and, 5.4 Auto adjustment analysis. The definition of the process of each run is based on the \texttt{actual\_parameters} module, where each selection is made by means of a combination of variables as logical values (\texttt{True or False}).
 
These variables are \texttt{sensitivity\_choice, multi\_run\_simulation, and auto\_adjust\_sensitivity\_test}.
 
Despite all these possibilities, the model's core is the same, but called to run in different ways.  

All destinations of outputs from SEAL model are identified by the four different simulation possibilities. Below is the scheme of logical variables and the resulted simulation that is run:

\begin{itemize}
	\item All logical variables are \texttt{False}: Single run
	\item Only \texttt{sensitivity\_choice}  is \texttt{True}: \texttt{Sensitivity\_analysis}
	\item Only \texttt{multi\_run\_simulation} is \texttt{True}:    \texttt{Multi\_Run\_Simulation}
	\item Only \texttt{auto\_adjust\_sensitivity\_test} is \texttt{True}: \texttt{Adjust\_test}
\end{itemize}

\subsection{Simple run or Test}
When the idea is to run a single simulation (a simple test), it is necessary to set all control logical variables (\texttt{sensitivity\_choice, multi\_run\_simulation, and auto\_adjust\_sensitivity\_test}) as \texttt{False}, and call the \texttt{main} module. Automatically, the SEAL model will run a simulation over the selected region and save results accordingly.  

The model will save results based on the selection of the saving option in the \texttt{actual\_parameters} module, so, check carefully the output saving options before running a simulation.

\subsection{Sensitivity analysis}

In order to run a \texttt{sensitivity\_choice}, select it as \texttt{True} on \texttt{actual\_parameters} and call the run from the module \texttt{control\_sensitivity\_analysis}. This module controls the multiple runs of the main module. As such, the model keeps working as a simple model, but called many times. 

The sensitivity analysis is performed over all parameters linked to the economics of the model (\texttt{ALPHA, BETA, QUANTITY\_TO\_CHANGE\_PRICES, MARKUP, LABOUR\_MARKET, SIZE\_MARKET, CONSUMPTION\_SATISFACTION, PERCENTAGE\_CHECK\_NEW\_LOCATION, TAX\_CONSUMPTION}), varying the values in between the limits of each one these parameters. For example, for \texttt{ALPHA} [0.01, 1]. 

Thus, \texttt{ALPHA} will be divided by the number of intervals defined by the user in the \texttt{control\_sensitivity\_analysis} module, set on the parameter \texttt{number\_of\_test\_parameter\_values} variable as an \texttt{integer} value. Using the same example of \texttt{ALPHA}, if it the parameter \texttt{number\_of\_test\_parameter\_values} is defined as 6, the values for \texttt{ALPHA} will be: 0.01;  0.208;  0.406;  0.604;  0.802;  1. 

However, note that, because of the number of parameters (9) and the example of 6 possible values for each one, a combinatory analysis of 10,077,696 possible combinations is generated. This would take a long time to run. Therefore, we chose to adopt a \texttt{ceteris paribus} instance in which each parameter new value is tested while the others remain fixed as “default”.

Therefore, when the user chooses the number of intervals, the state \texttt{True} for sensitivity analysis and runs the simulation calling \texttt{control\_sensitivity\_analysis.py}, the simulation will run all possible runs, as described, and produce a dataset of results, statistics and plots for each parameter. Analysis of the results may indicate how much the results vary based on each different set of parameter values. 

\subsection{Multi-run}

The \texttt{Multi-run} simulation is focused on identifying the consistence of the model, varying the random numbers involved on setting the decision-making processes. The idea is to replicate the same model, varying only the random numbers, so that we can isolate average consistent results, from outlier runs. We aim at reproducing a typical run along with pseudo variance information on its results. 

In order to run the \texttt{multi\_run\_simulation}, it is necessary to set the logical variable in \texttt{actual\_parameters} module to \texttt{True} and run the simulation calling the \texttt{control\_multi\_runs\_simulations.py}.
 
In this module, it is necessary to define the number of simulations to run \texttt{number\_of\_runs}. 

\subsection{Auto-adjustment analysis}

The option to use the Auto adjustment analysis aims at verifying the best combination of parameters for the model to produce optimal results. The process is based on three conditions: 

The subdivision of each parameter interval based on a number of possibilities defined by the user on \texttt{interval\_for\_values and the control\_autoadjust\_model} module. 
	
The pattern defined as a reference for the model to run the auto adjustment process. The parameters values that produce the maximum GDP and the minimum GINI index at same time in the last month of the simulation are set as default. 
	
The number of times the model will run in order to approximate the best values of parameters that lead to the optimal result is set on \texttt{control\_autoadjust\_model}, specifically setting the variable \texttt{times\_test\_aproximations}, as an integer number.
	
This structure was defined based on the possible number of combinations for the four most important economic parameters: \texttt{ALPHA, BETA, MARKUP, and TAX\_CONSUMPTION}.

In order to restrict a huge number of possibilities, a subdivision method for each parameter was adopted. Thus, in the beginning, the model divides the interval in a number of pieces. For example, for \texttt{ALPHA} and the \texttt{interval\_for\_values} set as 5, the values to be tested will be: 0.01,  0.2575,  0.505,  0.7525,  1 (the other three parameters will remain constant). Such a combination would produce a list of 255 possibilities of parameters. 

After the initial approximation, given the respective values of GDP and GINI, the best parameter interval is chosen and a subsequent division of the parameter values is applied. For example, if \texttt{ALPHA} was selected with values 0.505 and 0.7525 for the best combination of GDP and GINI, then the model will produce a new list of five values that lies within the previously set boundaries: 0.505, 0.57, 0.63, 0.69, 0.75; and so on until the number defined of \texttt{times\_test\_aproximations}. Using this approach, the total combinations of model process will be of 1,020 tests.

\subsection{Public Policy Test application: example}

A final alternative run is to set \texttt{alternative0} in the parameters module as \texttt{False}. In such a case, the model will run considering, for taxes purposes, all regions within each chosen Population Concentration Area (ACP) \cite{ibge._ministerio_do_planejamento_arranjos_2015} as a single region. It is possible to run \texttt{alternative0} and \texttt{multi\_run} simulations together. In such case, the runs alternate between \texttt{True} and \texttt{False alternative0}. 

A newer module \texttt{control\_multi\_ACPs\_alternate\_test} runs the \texttt{alternative0} test automatically and plots the differences and median of some indicators for both cases, municipalities as a single entity and municipalities as they are presently.

\section{Data input and requirements}\label{section 6}

All dataset used to run the model are available on the internet and come from official Brazilian Agencies websites, especially The Brazilian Institute of Geography and Statistics – IBGE.

The data used to run the model is fundamentally from the demographics Census of 2000 and 2010. The data used on the process are divided in Agents, Firms and Government. 

\subsection{Agents data from IBGE}

Agents data come from Demographics Census of 2000 (as a start point):

\begin{itemize}
	\item Tables of mortality probability by year, age group and gender. The data was interpolated using equal values to generalize it into every year data. Available on: \url{ ftp://ftp.ibge.gov.br/Projecao\_da\_Populacao/Projecao\_da\_Populacao\_2013/tabuas\_de\_mortalidade\_xls.zip}. 
	\item Tables of fecundity probability are available by year (between 2000 and 2030), group age (between 15 and 49 years old), and state. Available on: \url{
	ftp://ftp.ibge.gov.br/Projecao\_da\_Populacao/Projecao\_da\_Populacao\_2013/projecoes\_2013\_indicadores\_xls.zip}. 
	\item IDHM (Human Development Index) by municipalities – IPEA/IBGE, Census years (2000, 2010). This measure gives the development conditions for each municipality. Available on: \url{
	https://docs.google.com/gview?url=http://ivs.ipea.gov.br/ivs/data/rawData/atlasivs\_dadosbrutos\_pt.xlsx}
	\item Qualification by municipality and divided in years of study. Available on: \url{http://www.sidra.ibge.gov.br/bda/tabela/listabl.asp?z=cd&o=32&i=P&c=2986}
	
\end{itemize}

\subsection{Firms data from DATAVIVA}
\begin{itemize}
	\item Dataset for number of firms by municipalities, year (from 2002 to 2013), and productive sector. Provided by Ministry of Labor and Employment (MTE) and available on: 
	\url{http://dataviva.info/pt/data/}.
\end{itemize}

\subsection{Input spatial data}

\begin{itemize}
	\item The spatial dataset is also from IBGE. The model uses the municipalities boundaries maps, available on: 
	\url{ftp://geoftp.ibge.gov.br/organizacao\_do\_territorio/malhas\_territoriais/malhas\_municipais/municipio\_2010//}. 
	\item The urban area boundary definition uses the High Population Concentration Areas (ACPs), defined by IBGE for each metropolitan region of Brazil. These data can be found in: \url{ftp://geoftp.ibge.gov.br/cartas_e_mapas/mapas_do_brasil/sociedade_e_economia/areas_urbanizadas//areas_urbanizadas_do_Brasil_2005_shapes.zip}. These maps represent the spatial delimitation of urban areas (occurrence of populated areas). 
	\item The last demographics data are the urban and rural populations at each municipality, available on IBGE census website (table 200 on IBGE SIDRA database for 2000 and 2010): 
	\url{http://www.sidra.ibge.gov.br/bda/tabela/listabl.asp?z=cd&o=27&i=P&c=200}. 
\end{itemize}

\subsection{Instances generator}

The \texttt{generator} module is the one responsible for creating instances of all agents, families, regions, houses and firms. It follows official data and allocates them according to urban and rural proportion, and municipalities’ population numbers. 

Firstly, information on the proportion of urban inhabitants and number of firms by municipality are loaded. Then, information on 2000’s Municipal Human Development Index (HDM-I) is loaded. 

In order to create regions, the only necessary data is their HDM-I and their respective shapefile, read from 6.3. Such information is then passed on to class \texttt{Govern} in \texttt{Government}. 

All instances are saved in lists that are passed from \texttt{time\_iteration} to each process. Thus, they are dynamic lists with changing instances. 

For time saving purposes, agents for a given spatial configuration and percentage of population are saved in files that can later be loaded, configuring its persistence.

Using the newly created regions, the main function is \texttt{create\_all}, which creates, within each region, all other mentioned instances. 

The first step is to select, from population data for 2000, the number of inhabitants by age group and gender for each municipality. This procedure enables the successive creation of agents for each age group and, alternately, for each gender. 

Additionally, each agent gets: a qualification level assigned probabilistically, given municipal data of 2000; a random age that falls within his group age; a small random amount of money; and unique IDs. 

The number of families created in each region is proportional to the number of inhabitants with an average given by parameter \texttt{MEMBERS\_PER\_FAMILY} (typically set at 2.5 citizens per family).
 
Houses are above the number of families, given by parameter \texttt{HOUSE\_VACANCY} (typically set at 10\%). 

The number of firms, as stated before, are given by reading of actual data by municipality.  

The necessary additional functions are \texttt{create\_family, create\_household and create\_firm; allocate\_to\_family and allocate\_to\_households; and get\_random\_point\_in\_polygon}. 

\texttt{Create\_family} provides empty family vessels, which are then filled using \texttt{allocate\_to\_family}. For as long as there are agents to be allocated into families for a given municipality, one family and one agent are drawn from the available list and the match is made. There is one check to see if the agent does not belong to any family before allocation. That procedure ensures that the number of agents per family is random and the defined proportion is only valid for the total numbers per municipality. 

By construction, the number of families is always smaller than the number of houses available. Hence, all families that have a positive number of members are allocated randomly to an available household using \texttt{allocate\_to\_households}. In such a process – which of course happens only before the simulation actually begins – the family gets the ownership of that first house they are allocated to. The remaining houses that are empty are then subsequently distributed among all the families for a given municipality. 

That implies that, on average, 10\% of the families own their own house plus one other empty house that they may make available at the real estate market. Nothing withholds the possibility that luck will provide one family with two or three empty houses. 

The \texttt{create\_household} function is a bit more complex as it needs to abide to some spatial restrictions. Given the urban or rural proportion of the municipality, some of the houses will be set on rural areas (according to official urban municipal legislation) or on urban areas. That proportion is probabilistic and we do not guarantee that a specific fixed proportion will be urban or rural. 

The function that guarantees that actual address falls into the correct urban or rural areas of the map is \texttt{get\_random\_point\_in\_polygon}. This function uses \texttt{ogr.Geometry, geometry().Contains} and \texttt{AddPoint} within a given shapefile previously divided into urban and rural areas so that the allocation is correctly determined. 

Overall, besides address, the household is also given a random size drawn from the interval (20, 120), a fixed quality, from one to four and an endogenous price value given by \autoref{eq5}. Throughout the simulation, size and quality will remain the same, whereas I will be regularly updated, depending on the economic dynamics of the municipality. 

In the end, \texttt{create\_all returns} four lists: \texttt{my\_agents, my\_houses, my\_families and my\_firms}. \footnote{These objects are then saved using \texttt{pickles} so that in a subsequent run they can be uploaded and save running time}.

\subsection{Parameters and settings}

Parameters module contains not only all the parameters of the simulation, but also a lot of settings and decisions. We will describe them below, in order of appearance.
 
The first parameters is: \texttt{PERCENTAGE\_ACTUAL\_POP} and it establishes the percentage of the population that is going to be used in the simulation. A typical value is 0.01. However, the higher the proportion, the longest the time to complete the simulation. Simple tests can be run with 0.0001. 

The second parameter is a decision: \texttt{SIMPLIFY\_POP\_EVOLUTION}
and it accepts Boolean values (\texttt{True or False}). When \texttt{True}, the generation of agents is made using age groups. When \texttt{False}, the actual number of inhabitants by each specific age is used in the simulation. However, given smaller municipalities and smaller populations, it becomes inadequate to run a simulation of 1\% of the population of 54-year-olds, for example, when there are only 27 of those. That way, it is easier to reduce the population in bundles of age groups for small municipalities.

\texttt{LIST\_NEW\_AGE\_GROUPS} insures that the age groups are malleable. The modeler may define the superior limit of the groups. So far, we have been using [6, 12, 17, 25, 35, 45, 65, 100]. 

Next comes the decision on region of study. \texttt{processing\_states} can be any one of Brazilian 27 states or any combination of them, using the two capital letters of its name, such as 'MG' or 'SP'. To select them all, choose 'BR'.
 
In the sequence, it is possible to choose which Population Concentration Areas (ACPs), established by IBGE in 2005 within each state. ACPs approximate real conurbated metropolitan areas \cite{ibge._ministerio_do_planejamento_arranjos_2015}. Thus, they exclude rural, detached municipalities that may be officially part of metropolitan areas but that are not strongly integrated. 

Possible ACPs include: 

AM,	Manaus; PA	Bel\'em; AP	Macap\'a; MA	S\~ao Lu\'is, Teresina; PI	Teresina; CE Fortaleza, Juazeiro do Norte/Crato/ Barbalha; RN	Natal; PB	Jo\~ao Pessoa, Campina Grande; PE	Recife, Petrolina/Juazeiro; AL	Macei\'o; SE	Aracaju; BA	Salvador, Feira de Santana, Ilh\'eus/Itabuna, Petrolina/Juazeiro; MG	Belo Horizonte, Juiz de Fora, Ipatinga, Uberl\^andia; ES	Vit\'oria; RJ	Volta Redonda/Barra Mansa, Rio de Janeiro, Campos dos Goytacazes; SP	S\~ao Paulo, Campinas, Sorocaba, S\~ao José do Rio Preto, Santos, Jundiaí, S\~ao José dos Campos, Ribeir\~ao Preto; PR	Curitiba, Londrina, Maring\'a; SC	Joinville, Florian\'ipolis; RS	Porto Alegre, Novo Hamburgo/S\~ao Leopoldo, Caxias do Sul, Pelotas/Rio Grande; MS	Campo Grande; MT	Cuiab\'a; GO	Goi\^ania, Bras\'ilia; DF	Bras\'ilia

According to available data, a starting year has to be selected. For the case of Brazil, the starting data, based on Census data, is the year 2000.

\texttt{YEAR\_TO\_START}

The next selection is the folder where to storage output. It is essential for the running of the program and it has to be set in any new machine.

\texttt{OUTPUT\_PATH}

Then, a number of settings related to saving, output, plotting and type of simulation have to be completed with either \texttt{False or True}.

Another important choice is whether to run with actual municipalities or with merged/changed municipalities. \texttt{alternative0} is \texttt{True} indicates that boundaries are kept as is. When \texttt{False}, municipalities are merged together and collecting of taxes thus benefits all the citizens within a metropolitan region equally. 

\texttt{sensitivity\_choice} refers to the sensitivity analysis (section 5.2); the same is true for \texttt{multi\_run\_simulation} (section 5.3) and \texttt{auto\_ajdust\_sensitivity\_test} (section 5.4). As said before, a regular run should have all these parameters set to \texttt{False}.
 
\texttt{keep\_random\_seed} is necessary given the fact that \texttt{random} module is used in a number of different modules. In order to guarantee that all random numbers are drawn from the same sequence,  when needed, a \texttt{fixed\_seed = random.Random()} is in place.

Some of the settings/control are related to following the simulation as it unfolds, such as: \texttt{print\_statistics\_and\_results} and \texttt{show\_plots\_of\_each\_simulation}.
 
Another option is \texttt{time\_to\_be\_eliminated}, which is related to the amount of data that will enter the plots. The percentage determines the portion to be left out. For example, 0.2 will leave the first 20\% of the data out of the plots. 

Finally, a number of savings options: \texttt{save\_plots\_figures, save\_agents\_data, save\_agents\_data\_monthly, save\_agents\_data\_quarterly, save\_agents\_data\_annually, and create\_csv\_files}.

The module then initiates the actual simulation running parameters. Those include: \texttt{TOTAL\_DAYS}, measured in 21 working days a month, meaning that 5.040 days amount to 20 years.

\texttt{ALPHA}, which is the productivity factor-decaying exponent and can vary from 0 to 1, being 1 the most productive worker.

\texttt{BETA}, which is the consumption exponent and determines the amount the family decides to consume/save. The larger the exponent, more consumption and less savings. 

\texttt{QUANTITY\_TO\_CHANGE\_PRICES}, which is the number threshold that firms check against their stocks in order to decide whether demand is high (below the quantity) and thus prices should rise. 

\texttt{MARKUP}, which is the amount the firm applies when varying prices. 

\texttt{LABOUR\_MARKET}, which is the frequency that the firm makes decisions about entering the market. The higher the parameter, less often. 

\texttt{SIZE\_MARKET}, the number of firms that the consumer check for prices and distances when deciding where to shop.

\texttt{CONSUMPTION\_SATISFACTION}, this is just a parameter that scales down consumption in order to keep track of accumulated satisfaction given by consumption.

\texttt{PERCENTAGE\_CHECK\_NEW\_LOCATION}, the percentage of total families that enter the real estate market each month. 

\texttt{TAX\_CONSUMPTION}, the general consumption tax applied to all sales and collected by municipal government. 

The process of rearrange population data is performed after the definition of parameters values. Can be chose to perform a simplification of population when the percentage of population chosen are too small. In the end the \texttt{parameters} module save the files for control process and generate the output \texttt{*.txt} data.

\subsection{Necessary \texttt{Python} libraries}

We run the model in \texttt{Python 3.4.4}.

The following \texttt{Python} libraries are necessary:

\begin{verbatim}
from numpy import median
from operator 
    import methodcaller, attrgetter
from osgeo import ogr
from pandas import read_csv
from timeit import default_timer
ggplot
glob
itertools
matplotlib.pyplot
numpy
os
pandas
pickle
random
subprocess
sys
\end{verbatim}

A simpler way to install the necessary environment is: 

\begin{verbatim}
Install 
Anaconda3-2.3.0-Windows-x86_64
\end{verbatim}

Then type in \texttt{Terminal}
\begin{verbatim}
conda install python=3.4.4
conda install pandas=0.18.0
conda install numpy=1.10.4
conda install -c ioos 
    geopandas=0.2.0.dev0
pip install ggplot==0.6.8
\end{verbatim}
\section{Outputs}\label{section 7}

\texttt{Output} module is called at the end of the month and it is divided into general statistics \texttt{call\_statistics} and regional statistics \texttt{call\_regional\_stats}. The former calls \texttt{my\_statistic}s and writes \texttt{TXT} outputs. The information it contains include: 

\begin{verbatim}
actual_month
price_index
GDP_index
unemployment
average_workers
families_wealth
firms_wealth
firms_profit
GINI_index
average_utility
\end{verbatim}

The regional statistics include: 

\begin{verbatim}
actual_month
region_id
commuting 
get_pop
get_gdp 
regional_gini
regional_unemployment
index
GDP_region_capita
\end{verbatim}

Some of the methods are detailed in the next subsection. Finally, output module also includes \texttt{save\_agents\_data} and \texttt{sum\_region\_gdp} functions.

\subsection{Statistics}

The \texttt{statistics} class is just a bundle of functions together without a permanent instance of data. Thus, every time \texttt{Statistics()} is called, it is initiated anew. The functions include average price of the firms, regional GDP, based on FIRMS' revenues, GDP per capita, unemployment, families' wealth, GINI, regional GINI and commuting information.

\texttt{average\_price} method checks for every firm in \texttt{my\_firms} the quantity and price of each product (currently just one). Then it calculates and returns the average. 

\texttt{calculate\_region\_GDP} goes through the firms in a given region and adds up the cumulative sold value, inclusive of taxes.
 
\texttt{update\_GDP\_capita} also goes through the firms in the region, adding up cumulative value sold and then dividing it by current population.

\texttt{calculate\_unemployment} goes through all agents checking whether they are minor or retired and then classifying those left as employed or unemployed. Then it updates unemployment rate as the ration between unemployed and the total workforce (employed plus unemployed).

\texttt{calculate\_families\_wealth} returns the sum of all families wealth. 

\texttt{calculate\_utility} returns the average of families' average utility. That is, first, the average within each family is calculated and then we calculate the average of such values. 

\texttt{calculate\_GINI} is given by the typical GINI coefficient calculated upon this very families’ average utility. That is, the GINI is in fact the inequality present among families cumulative consumption. \texttt{calculate\_regional\_GINI} is made exactly the same way but considering only the families residing in a given region (municipality).

\texttt{update\_commuting} is calculated for every member of each family in each region that \texttt{is\_employed}. 

\subsection{Plotting and analysis}

The model produces some data and graphs by default in the \texttt{plotting} module and others in accordance to users' choices. The plotting of the model may use different modules.

In all models, the plotting module produces the same set of plots from the \texttt{temp\_general\_\%s.txt} data. This output data represents the aggregated measures of all regions in the simulation. These measures are:  \texttt{Price index, GDP, Unemployment, Average of number of workers by firms, Families wealth, Firms wealth, Firms profit, GINI index, Average utility}. For each one of these measures one plot is produced to allow for a general understanding of the patterns observed in the running model. 

The same module produces the firms' plots using data from \texttt{temp\_firm\%s.txt}. They represent the \texttt{Amount produced, Price, Mean and Median of Number employees}. They are used to understand the firms' general behavior.

Finally, the \texttt{plotting} module produces regional stats (one value for each measure in each municipality). The measures are: \texttt{Commuting, Population, GDP of region, GINI by region, Unemployment by municipality, Quality life index by municipality, GDP per capita by municipality}.

When the model is in a Multi Run simulation mode, the module \texttt{plot\_multi\_run.py} use the values of \texttt{temp\_general\_\%s.txt} and stack up each measure (i.e. Unemployment) of all simulations in the same graph. The same process is observed for regional data and agents' data. They are used to produce plots considering each municipality. 

\subsection{Output data}

Output data represents the minimal data necessary to analyze the results of the model. Therefore, all values of simulation for general and regional data are saved by default. These data contain the economics and social measures for the whole selected region \texttt{temp\_general\_\%s.txt} and the regional measures (by municipality) of the social and economic measures \texttt{temp\_regional\_stats\%s.txt}.
 
The user can choose the scale of agents' data in the files: \texttt{temp\_firm\%s.txt}, \texttt{temp\_agent\%s.txt} and \texttt{temp\_house\%s.txt}, and the time frequency to save such data: monthly: \texttt{save\_agents\_data\_monthly}, quarterly, \texttt{save\_agents\_data\_quarterly} or annually, \texttt{save\_agents\_data\_annually}. Such information is used to produce aggregated or simple plots (in accordance to the type of model running). 

\subsubsection{Details of TXTs files}

When the option to save agents data is set to \texttt{True}, all data is saved and 5 \texttt{*.TXT} files are created. 

The files are: `agent, firm, general, house` and `regional`. Family information is also available in the house file.
They contain no headers, delimiter = , and decimal = .

$AGENT$ contain the following columns:
\texttt{month	region\_id	gender	long	lat	id	age	qualification	firm\_id	family\_id	utility	distance}

$FIRM$ contain the following columns:
\texttt{month	firm\_id	region\_id	long	lat	total\_balance\$	number\_employees	total\_quantity\_in\_stock	amount\_produced	price}

$GENERAL$ contain the following columns:
\texttt{month	price\_index	gdp\_index	unemployment	average\_workers	families\_wealth	families\_savings	firms\_wealth
firms\_profit	gini\_index	average\_utility}

$HOUSE$ contain the following columns:
\texttt{month	house\_id	long	lat	house\_size	house\_price	family\_id	family\_savings	region\_id}

$REGIONAL$ contain the following columns:
\texttt{month	region\_id	commuting	pop	gdp\_region	regional\_gini	regional\_unemployment	qli\_index	gdp\_percapta
treasure}

The file's names always include the following parameters: \texttt{"None", 	alternative0, PERIODICITY\_SAVE\_DATA, TOTAL\_DAYS, total\_pop, SIZE\_MARKET, ALPHA, BETA, QUANTITY\_TO\_CHANGE\_PRICES, MARKUP, LABOUR\_MARKET, CONSUMPTION\_SATISFACTION, PERCENTAGE\_CHECK\_NEW\_LOCATION}, and \texttt{TAX\_CONSUMPTION}. 

\section{Design concepts}\label{section 8}

There are two main drives of the agent-based model here presented. Firstly, it is a restricted adherence to investigate public policies within a spatial continuum. In order to achieve that, there are locational Cartesian attributes for houses and for firms with agents commuting the distance. Further, region space – organized in municipalities contain both their actual geodesic boundaries, which determines their action space, but also comprehend the urban rural divide, which enables urban concentration and rural sparseness. Thus, distance explicitly is considered within interaction in the labor market and consumption market. Further, there is a real estate market that is very much dependent on how well the firms in the region are performing. Stronger firms with higher sales increase taxes, which are directly applied to improving estates quality and directly, prices. 

Secondly, the model is proposed in a rather flexible and generic way so that it can be constituted as a framework for later public policy analysis. Agents, firms, markets, space, activities are constructed so that newer research questions can be easily adapted into the model, which would provide rapid answering. As it is, for instance, results can be provide for an alteration in a specific tax or a change in a firm’s strategy to enter the labor market, or a municipality fusion or, yet, in the influence of those changes in mobility patterns, for instance. 

Apart from those two specific design concepts, obviously, the model follows the typical agent-based specifications, that is: agent heterogeneity, decision-making based on local variables, and emergence.

Specifically, we can say that the agents in this model adapt in the sense that they change behavior giving observed variables at their environment. For example, firms change decision-making (and salaries paid) depending on whether they have available balance and profits. Agents decide whether to move into a better quality house depending on the proportion of employed adults in the family. 

The model is not intended as a forecasting tool at this stage. The focus is more on setting a framework, learning about complex interactions among different factors across a large spectrum and gaining magnitude of changes insights. 

Interactions are plenty. Specifically, they happen in the three proposed markets and within the families (co-responsible for consumption, savings and utility). Indirectly, there is interaction among those agents within the same region. 

Stochasticity is present in the model with the use of drawing and random decision-making in a number of procedures. However, a control variable is present that enables the reproduction of the same result. Typically, results should be presented as summaries of a number of simulated runs. 

A collective entity that is relevant for the model is that of the family (see section 4.2). 

\section{Final Considerations}\label{section 9}

This is still a model in progress. However, at this stage, we believe we have most of the framework as intended. 

Thus, this research project reveals from a code perspective nearly all aspects of the model. That includes a description of purpose and concept, model scheduling, detailed information and methods on all agents and classes used, the market interactions, data generated and output produced. It also includes information on running the model for multiple runs and explicitly using it as a policy example (section 5.5).

However, we have yet to publish major research studies using such framework, except for \cite{furtado_simple_2016, furtado_da_2016}. This is exactly the next step: apply the framework to public policy research questions. 

Finally, we would like to add that the current working team is short and we would be willing to cooperate with other fellow scientists. 

\section*{Acknowledgment}
The authors would like to thank Institute for Applied Economic Research and especially the Department of Innovation and Infrastructure (DISET) for their continuing support. 

\bibliography{seal_arxiv}
\bibliographystyle{IEEEtran}

\end{document}